\documentclass{aa}

\usepackage{chemmacros}

\newcommand{\ha}{H$\alpha$} 
\newcommand{\hb}{H$\beta$}

\newcommand{\helium}{He\,{\sc i}}

\newcommand{\nia}{[N\,{\sc i}]~5199~\AA}
\newcommand{\oi}{[O\,{\sc i}]~6300~\AA}

\newcommand{\niia}{[N\,{\sc ii}]~5755~\AA}
\newcommand{\niib}{[N\,{\sc ii}]~6584~\AA}
 
\newcommand{\siia}{[S\,{\sc ii}]~6717~\AA}
\newcommand{\siib}{[S\,{\sc ii}]~6731~\AA}

\newcommand{\oiiia}{[O\,{\sc iii}]~4959~\AA}

\newcommand{\nitrogen}{[N\,{\sc ii}]}

\newcommand{\nitrogena}{[N\,{\sc i}]}
\newcommand{\oxygeniii}{[O\,{\sc iii}]}

\newcommand{\oxygeni}{[O\,{\sc i}]}

\newcommand{\sulfur}{[S\,{\sc iii}]}
\newcommand{\sulfurt}{[S\,{\sc ii}]}

\newcommand{\ironii}{[Fe\,{\sc ii}]}
\newcommand{\ironiii}{[Fe\,{\sc iii}]}

\newcommand{\carboni}{[C\,{\sc i}]}
\newcommand{\carboniia}{C\,{\sc ii}}

\def\vhel{\ifmmode{V_{{\rm HEL}}}\else{$V_{{\rm HEL}}$}\fi}
\def\vsys{\ifmmode{V_{\rm sys}}\else{$V_{\rm sys}$}\fi}
\def\kms{\ifmmode{~{\rm km\,s}^{-1}}\else{~km~s$^{-1}$}\fi}
\def\vlsr{\ifmmode{v_{\rm lsr}}\else{$v_{\rm lsr}$}\fi}

\usepackage{graphicx}
\usepackage{amsmath}    % Advanced maths commands
\usepackage{amssymb}    % Extra maths symbols
\usepackage{txfonts}
\usepackage[T1]{fontenc}
\usepackage{ae,aecompl}

\usepackage{newtxtext,newtxmath}
\usepackage{pdflscape}
\usepackage{longtable}
\usepackage{supertabular}
\usepackage{float}

\usepackage{rotating}
\usepackage{footnote}
\usepackage{times}

\usepackage[colorlinks=true, urlcolor=blue, linkcolor=blue, citecolor=blue]{hyperref} 
\usepackage{graphicx}

\definecolor{lime}{HTML}{A6CE39}
\DeclareRobustCommand{\orcidicon}{
        \begin{tikzpicture}
        \draw[lime, fill=lime] (0,0) 
        circle [radius=0.16] 
        node[white] {{\fontfamily{qag}\selectfont \tiny ID}};
        \draw[white, fill=white] (-0.0625,0.095) 
        circle [radius=0.007];
        \end{tikzpicture}
        \hspace{-2mm}
}

\foreach \x in {A, ..., Z}{\expandafter\xdef\csname orcid\x\endcsname{\noexpand\href{https://orcid.org/\csname orcidauthor\x\endcsname}
                        {\noexpand\orcidicon}}
}

\begin{document}

   \title{Detection of the \carboni~$\lambda$8727~emission line}

   \subtitle{Low-ionization structures in NGC~7009}

   \author{S. Akras\inst{1}{\orcidA{}}, H. Monteiro\inst{2}{\orcidB{}}, J. R. Walsh\inst{3}{\orcidC{}}, L. Konstantinou\inst{1}{\orcidD{}}, D.R. Gon\c{c}alves\inst{4}, J. Garcia--Rojas\inst{5,6}{\orcidE{}}, P.~ Boumis\inst{1}{\orcidG{}} and I. Aleman\inst{2}{\orcidF{}}}

    \authorrunning{S. Akras et al.}

    \institute{Institute for Astronomy, Astrophysics, Space Applications and Remote Sensing, National Observatory of Athens, GR 15236 Penteli, Greece,
        \email{stavrosakras@gmail.com}
         \and
             Instituto de F\'isica e Qu\'imica, Universidade Federal de Itajub\'a, Av. BPS 1303, Pinheirinho, 37500-903, Itajub\'a, MG, Brazil 
         \and
             European Southern Observatory. Karl-Schwarzschild Strasse 2, D-85748 Garching, Germany
         \and
             Observat\'orio do Valongo, Universidade Federal do Rio de Janeiro, Ladeira Pedro Antonio 43, 20080-090, Rio de Janeiro, Brazil
         \and 
             Instituto de Astrof\'isica de Canarias, E-38205 La Laguna, Tenerife, Spain
         \and
             Departamento de Astrof\'isica, Universidad de La Laguna, E-38206 La Laguna, Tenerife, Spain
         }
   \date{}

% \abstract{}{}{}{}{} 
% 5 {} token are mandatory
 
  \abstract
    { We report the first spatially resolved detection of the near-infrared \carboni~$\lambda$8727 emission from the outer pair of low-ionization structures (LISs) in the planetary nebula NGC~7009 from data obtained by the Multi Unit Spectroscopic Explorer (MUSE) integral field unit. This atomic carbon emission marks the transition zone between ionized and neutral gas and for the first time offers direct evidence that LISs are photodominated regions. The outer LIS pair exhibits intense \carboni~$\lambda$8727 emission, but \helium$~\lambda$8733 is absent. Conversely, the inner pair of knots shows both lines, likely due to the host nebula emission. Furthermore, the \carboni~$\lambda$8727 line is absent in the host nebula emission, but \helium~$\lambda$8733 is present. Although the origin of the \carboni~$\lambda$8727 line is still debated, its detection supports the scenario of photoevaporated dense molecular clumps. 
    }

\maketitle
%
%-------------------------------------------------------------------
%
\section{Introduction}

The origin of microstructures, such as knots, filaments, or jet-like features with a strong \nitrogen~$\lambda$6584 emission line, that have been reported for a number of planetary nebulae (PNe) by \citet[][]{Balick1987} and it has been a long-standing problem in the PNe formation as well as stellar evolution. \cite{Corradi1996} demonstrated based on \ha+\nitrogen/\oxygeniii\ CCD imaging that the low-ionization microstructures are far more common in PNe than we knew before. Based on their kinematic characteristics and expansion velocities, these structures occur in two main groups: (i) fast low-ionization emission regions \citep[FLIERs, ][]{Balick1993}, and (ii) slow-moving low-ionization emitting regions \citep[SLOWERs, ][]{Perinotto2000}. In 2001, \cite{Goncalves2001} integrated both classes into a more general category, called low-ionization structures (LISs), and conducted a thorough comparison between formation models and observations. When LISs are found in pairs, they are also known as ansae, that is, the knots at the tip of jet-like features along the major axis of the nebulae that are located at a nearly equal distance from the central star \cite[e.g. NGC~3242, NGC~7009, Hb~4, ][]{Balick1987,Soker1990,GarciaSegura1997}.

Numerous studies of LISs have been carried out since they were first reported, either via spectroscopy or imaging \citep[e.g. ][among others]{Balick1993,Balick1994,Hajian1997,Balick1998,Corradi1999,Corradi2000a,Corradi2000b,Goncalves2003,Goncalves2004,Goncalves2006,Goncalves2009,Akras2016,Ali2017,Balick2020,Akras2022,Mari2023a,Mari2023b}. The interpretation of the enhanced \nitrogen~emission from LISs relative to the surrounding nebular gas as a result of an overabundance of nitrogen \citep{Balick1994} has been ruled out because most PNe present no difference in chemical composition between the LISs and the host PNe \citep[e.g. ][]{Hajian1997,Goncalves2006,Akras2016,Ali2017,Akras2022,Mari2023a,Mari2023b}. 

The intense ultraviolet (UV) radiation from the central stars of PNe and shocks between the nebular components or the interstellar medium have been suggested to explain the observed enhancement of low-ionization lines such as \nitrogen~$\lambda\lambda$6548,6584, \sulfurt~$\lambda\lambda$6716,6731, and \oxygeni~$\lambda$6300, which originate from LISs \citep[e.g.][]{Dopita1997,Hajian1997,Goncalves2003,Goncalves2004,Akras2016,Ali2017,Mari2023a,Mari2023b}. No definitive conclusion has been reached so far. %Overall, neither photo-ionization nor shocks can adequately reproduce all the physical properties and observed emission line ratios. 

The main difference between the LISs and the surrounding host nebulae is the electron density of the gas (n$_{\rm e}$). In particular, n$_{\rm e}$ is systematically lower in LISs than in the nebular gas \citep[e.g. ][]{Mari2023b}. The detection of molecular hydrogen (H$_2$) emission in LISs \citep{Akras2017,Fang2018,Akras2020a} has unveiled an additional mass component, as proposed by \citep{Goncalves2009}. Hence, LISs are most likely dense molecular clumps, in agreement with various formation models \citep[e.g.][]{Steffen2001,Raga2008,Balick2020}. 

This molecular component has only been directly associated with the cometary knots of the Helix \citep{Matsuura2009,Andriantsaralaza2020}, Dumbbell \citep{Kwok2008}, and Ring nebulae \citep{Speck2003}. Manchado and colleagues unveiled through adaptive optics imaging that the H$_2$ emission that was previously detected in the equatorial region of the bipolar PN NGC~2346 \citep{Arias2001} is fragmented into clumps and filaments \citep{Manchado2015}. More recently, the astonishing images from the {\it James Webb} Space Telescope (JWST) revealed the fragmentation of H$_2$ emission into knots and filaments in NGC~3132 \citep{Demarco2022} and NGC~6720 \citep{Wesson2024}.

Low-ionization structures have a typical size of 10$^{16}$-10$^{17}$~cm and are found in young PNe $<$2000~yr, and the cometary knots are larger and are found in older PNe $>$7000~yr \citep{Akras2020c}. Both are characterized by high core densities (n$_H>$10$^5$~cm$^{-3}$) that shield the H$_2$ gas against the photodissociating UV radiation \citep[e.g][]{ODell2005,ODell2007}.

The pair of outer LISs in the archetypal PN NGC~7009 is of particular interest. The analysis of narrow-band near-IR imaging data has revealed high H$_2$ 1-0/2-1 and moderate H$_2$ 1-0/Br$\gamma$ ratios, \citep[see fig.~6 in ][]{Akras2020c} which are both consistent with photoexcited and shock-excited gas. In addition to the H$_2$ emission, \ironii~$\lambda$1.644~$\mu$m emission has also been found to emanate from the same pair of LISs in NGC~7009 \citep{akras2024accepted}. The scenario of photoevaporating clumps is favored due to the spatial distribution and line stratification \citep{Akras2022}.

In this paper, we add one more piece of evidence to solve the puzzling problem of LISs in PNe. We report the first detection of the near-infrared \carboni~$\lambda$8727 emission line that is directly associated with the LISs in NGC~7009. In Sect.~2, we discuss the previous detections of the \carboni~$\lambda$8727 line in PNe. The results of our analysis of NGC~7009 are described in Sect.~3. In Sect.~4, we discuss the origin of the \carboni~$\lambda$8727 emission, and we conclude in Sect.~5.

\section{Near-infrared \carboni~emission lines in planetary and other nebulae}

The emission lines from neutral carbon in the near-infrared (NIR, 750-3000~nm) such as \carboni~$\lambda$8727 (5->4) and $\lambda\lambda$9824 (4->2), 9850 (4->3) were first detected in the highly ionized planetary nebula NGC~7027 by \cite{Danziger1973} in 1973. NGC~7027 is a C-rich nebula with several carbon-based molecules such as CO, CH$^+$, and C$_2$H, which were found in spectra recorded with the photodetector Array Camera and
Spectrometer (PACS) and Spectral and Photometric Imaging REceiver (SPIRE) \citep{Wesson_etal_2010}. The neutral \carboni~emission lines likely originate from a dense and warm region in which CO is partially dissociated. The \carboni~emission lines have also been detected in the far-infrared centered at 370.4 and 609.14 $\mu$m with Herschel, but the emission was not spatially resolved (NGC~7027; \citealt{Wesson_etal_2010} or NGC~6781; \citealt{Ueta2014}).

\cite{Jewitt1983} carried out a \carboni~survey of PNe and found that three (NGC~6210, NGC~6720, and NGC~7027) out of six PNe display \carboni~emission lines in their spectra. The detection in NGC~6720 is particularly interesting. The \carboni~$\lambda\lambda$9824+9850 emission lines were found to be cospatial with the \oxygeni~$\lambda$6300 line. All these emission lines originate from dense filaments distributed in the nebular shell. Collisions of carbons with free electrons were proposed to be the cause for the neutral carbon emission lines in NGC~6720. \cite{Jewitt1983} stated that the \carboni~$\lambda$8727 emission line likely originates from regions of the PNe that are shielded from the UV radiation from the central nucleus. The spectroscopic study of NGC~7027 and NGC~6720 by \cite{Liu1996} confirmed the previous detection of the forbidden \carboni~lines. The same authors also reported a cospatial distribution of the \carboni~$\lambda$8727 and \nitrogena~$\lambda$5200~lines for NGC~6720, suggesting that both lines emanate from the same partially ionized regions. Interestingly, the electron temperature computed from the \carboni~lines ($T_{\rm e}$\carboni) was found to be systematically lower by 2000-3000~K than the temperatures obtained from the traditionally used diagnostic lines of \nitrogen, \oxygeniii, and \sulfurt. This result is consistent with the decrease of $T_{\rm e}$ in the denser regions of gas \citep[][]{Liu1995CI,Liu1996}.

\cite{Liu1995CI} reported the detection of the \carboni~$\lambda$8727 and $\lambda\lambda$9824, 9850 lines in four additional PNe (NGC~2346, NGC~2440, NGC~3132, and IC~4406). Notably, molecular hydrogen lines have been reported for all four PNe \citep[e.g][]{storey1984,kastner1996,Hora1999,Arias2001,Manchado2015,Mata2016,Demarco2022}.

\begin{table}[h!]
\centering
\caption{Planetary nebulae in which the \carboni~$\lambda$8727 and/or $\lambda\lambda$9824,9850 emission lines have been detected.}
\label{tab:CI_detections}
\begin{tabular}{lccl}

\hline                                  
PN      &       \carboni & \carboni & Refs.     \\
name   &  $\lambda$8727    &  $\lambda\lambda$9824,9850    \\
\hline          
NGC~7027 &      \checkmark      &        \checkmark & 1, 2 \\
NGC~6210 &      \checkmark      &        \checkmark & 2\\
NGC~6720 &      \checkmark      &        \checkmark & 2 \\
NGC~2346 &      \checkmark      &        \checkmark & 3 \\
NGC~2440 &      \checkmark      &        \checkmark & 3 \\
NGC~3132 &      \checkmark      &        \checkmark &  3\\
IC~4406  & \checkmark & \checkmark & 3 \\
NGC~6741 &      \checkmark      &       \checkmark  & 4         \\
NGC~7662 &      \checkmark      &       - & 5\\
IRAS 1385-5517  & \checkmark &  - & 6 \\
NGC~6543 &      -       &       \checkmark & 7 \\
IC~5117  & \checkmark$^{\dag\dag}$ & \checkmark$^{\dag\dag}$ & 8 \\
IC~418  & \checkmark & \checkmark & 9, 17\\
He~2-138  & \checkmark & - & 10 \\
BoBn-1  & \checkmark &  - & 11 \\
NGC~6369 &      \checkmark  &   - &  12\\
M~1-25  &       \checkmark      & - & 12        \\
M~1-30  &       \checkmark      &       - & 12 \\
He~2-86  & \checkmark & - & 12  \\
M~1-61  & \checkmark & -& 12 \\
Cn~1-5  & \checkmark & -& 12\\
Hb~4  & \checkmark &    - &12 \\
PC 14  & \checkmark & - & 12\\
Pe 1-1  & \checkmark & - & 12   \\

NGC~6302 &      \checkmark      &        - & 13 \\
NGC~3918 &      \checkmark      &       \checkmark & 14 \\
Lin49  & \checkmark &   \checkmark & 15 \\
Hen 3-1357 & \checkmark &       - & 16 \\
NGC~5315 &      \checkmark      &        \checkmark & 18 \\
M~1-31  &       \checkmark      &       \checkmark & 19  \\
M~1-33  &       \checkmark      &       \checkmark & 19 \\
M~1-60  &  \checkmark  & \checkmark & 19 \\
M~2-31  & \checkmark & \checkmark & 19 \\
Hen 2-73  & \checkmark &        \checkmark & 19 \\
Hen 2-96  & \checkmark &        \checkmark & 19 \\
H~1-40  & \checkmark &  \checkmark & 19\\
H~1-50  & \checkmark &  \checkmark & 19 \\
J~900  & \checkmark &   - & 20 \\ 
H~4-1   & \checkmark?$^{\dag}$ &        \checkmark & 21 \\
\hline
\end{tabular}
\tablefoot{
\dag~An emission line centered at 8727~\AA~was detected, but was identified as \carboniia. $^{\dag\dag}$ An 1~\AA~difference is reported between the $\lambda$(obs) and $\lambda$(lab). 
}
\tablebib{
(1)~\citet{Danziger1973}; (2) \citet{Jewitt1983}; (3) \citet{Liu1995CI}; (4) \citet{Hyung_1997}; (5) \citet{Hyung_1997b}; (6) \citet{Sivarani1999}; (7) \citet{Hyung_2000}; (8) \citet{Hyung_2001}; (9) \citet{Sharpee2004}; (10) \citet{Williams2008}; (11) \citet{Otsuka_2010}; (12) \citet{Garcia_Rojas_2012}; (13) \citet{Rauber2014}; (14) \citet{Garcia_Rojas_2015}; (15) \citet{Otsuka_2016}; (16) \citet{Otsuka_2017}; (17) \citet{Dopita_2017}; (18) \citet{Madonna_2017}; (19) \citet{Garcia_Rojas_2018}; (20) \citet{Otsuka_2019}; (21) \citet{Otsuka_2023}.
}
\end{table}

Three possible mechanism were discussed as the cause of the emission of the NIR forbidden \carboni~lines by \cite{Liu1995CI}: (i) collisional excitation by electrons \citep{Danziger1973,Jewitt1983}, (ii) recombination of C$^+$ ions in a dense gas illuminated by an intense UV field \citep{Escalante1991}, and (iii) UV  continuum fluorescence emission. The three mechanisms can be distinguished through the \carboni~line ratio \citep[R(\carboni)=($\lambda\lambda$9824+9850)/$\lambda$8727, see fig.~2 in][]{Liu1995CI}. R(\carboni)$>$10 implies a collisional excitation origin, while a ratio between 5 and 10, depending on $T_{\rm e}$, can be attributed to both collisional excitation and recombination processes. Collisional deexcitation may also have an impact on the line intensities of high-density regions (critical densities for the \carboni~$\lambda$8727, $\lambda$9824, and $\lambda$9850 lines for T$_{\rm e}$=1000~K are $\sim$1.3$\times$10$^7$,$\sim$1.4$\times$10$^4$, and $\sim$1.4$\times$10$^4$ cm$^{-3}$, respectively). The pure UV fluorescence process yields line ratios close to the lower recombination bounds ($\sim$5), but significantly weaker emission line intensities \citep{Liu1995CI}. Additionally, Liu et al. found evidence that collisional excitation by electrons was the dominant mechanism that excited the \carboni\ lines in their four PNe.

\begin{figure}
\centering
\includegraphics[width=8cm]{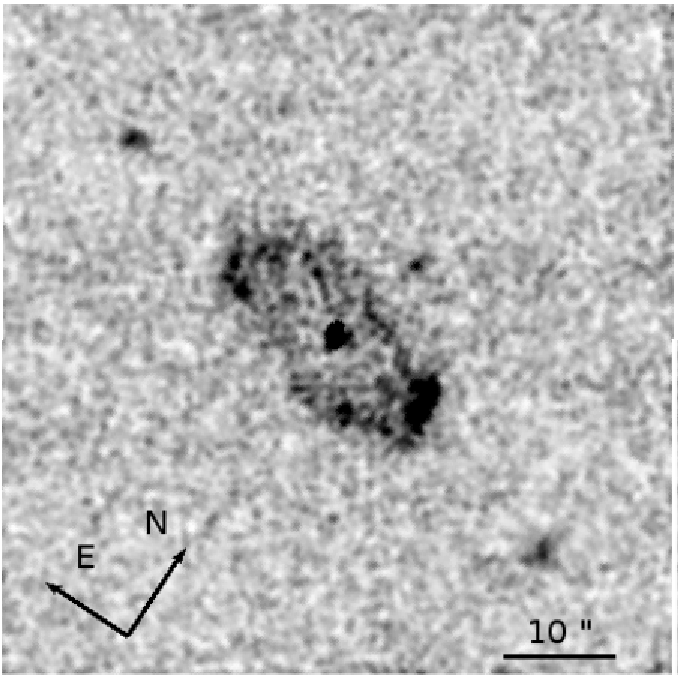}
\caption{ \carboni~$\lambda$8727 emission line in the MUSE map of NGC~7009 in 1x1 spatial binning.}
\label{figCI}
\end{figure}

The \carboni~$\lambda$8727 and/or $\lambda\lambda$9824,9850 emission lines, depending on the wavelength coverage of the spectrograph, were also detected in another 39 PNe that are listed in Table~\ref{tab:CI_detections}. The detection of the lines in the post-AGB star IRAS 11385-5517 indicates a nebula with a very low degree of ionization. All the aforementioned \carboni~detections were made through long-slit spectroscopy, and the spatial extent of the regions that emitted these lines is still unknown. 
Interestingly, \cite{GarciaRojas2022} presented similar spatially resolved \carboni~$\lambda$8727 MUSE maps for the PNe NGC~6778 and M~1-42. Both PNe display a knotty structure in the NIR forbidden \carboni~$\lambda$8727 line, which appears to be cospatial with the \oxygeni~$\lambda$6300 and \nitrogena~$\lambda$5200 lines. An additional pair of LISs in the PN NGC~3242 was found to emit the \carboni~$\lambda$8727 line (Konstantinou et al. in preparation). It is truly intriguing that the \carboni~$\lambda$8727 line has been found in a few LISs.

In addition to PNe, the same NIR neutral carbon lines were detected in the photodissociation regions (PDRs) of H~{\sc ii} regions, such as NGC~2024 \citep{Munch1982}, NGC~2023 \citep{Burton1992}, the Orion nebula \citep{Munch1982, Weilbacher2015}, 30~Dor \citep{castro2018}, M~17, and NGC~346 \citep{Henney2024}. \cite{Burton1992} argued for NGC 2023 that the \carboni~$\lambda$8727 and H$_2$ lines originate from the same high-density ($n_{\rm H}>$10$^{5}$ cm$^{-3}$) gas in a PDR. \cite{Stock2011} reported the detection of the NIR carbon lines in the NGC~3199 nebula around the WR~18 Wolf-Rayet star. These authors argued that the recombination process for C ions is more likely than collisional excitation to explain the observed intensities of the lines.

Overall, \carboni~is considered as a good tracer for H$_2$ regions because the same energy photons can ionize C$^0$ and O$^0$ and dissociate H$_2$. PDR modelling can naturally explain the common origin of the H$_2$ \oxygeni~and \carboni~lines from a dense partially ionized and/or neutral gas with an optical depth Av$<$2 \citep{Tielens1985a,Tielens1985b,Hollenbach1999}. Although a recombination of C$^+$ ions is a plausible mechanism for the origin of the NIR neutral carbon lines in PDRs  \citep{Escalante1991,Munch1982,Burton1992}, collisional excitation in a partially ionized gas is more likely in PNe \citep{Danziger1973,Liu1995CI,Liu1996}. The difference in the excitation mechanisms of \carboni~lines in various environments is attributed to the strength of the UV radiation field and to the physical properties of the gas. 

Moreover, the \carboni~($\lambda\lambda$9824+9850)/$\lambda$8727 line ratio can also be used as a temperature diagnostic for neutral and partially ionized gas, similar to the \oxygeni~diagnostic lines ($\lambda\lambda$6300+6363)/$\lambda$5577).

\begin{table}
\caption{Log of the MUSE observations of NGC\,7009 in WFM-NOAO-E mode.}
\label{Tab:Obs}
\begin{tabular}{clcll}
\hline
UT Start & \# & Exp. time  &  Airm. & Seeing \\
& & (s) &  & ($"$) \\\hline
 2016-07-07 09:05:01.393&1/9&10&1.235&1.14\\
 2016-07-07 09:07:10.071&2/9&30&1.242&0.9\\
 2016-07-07 09:09:38.672&3/9&150&1.251&0.89\\
 2016-07-07 09:13:33.127&4/9&120$^{\rm a}$&1.264&1.06\\
 2016-07-07 09:17:37.157&5/9&150&1.282&0.92\\
 2016-07-07 09:22:05.227&6/9&150&1.3&0.92\\
 2016-07-07 09:25:59.464&7/9&120$^{\rm a}$&1.315&0.96\\
 2016-07-07 09:30:02.676&8/9&150&1.335&0.91\\
 2016-07-07 09:34:51.104&9/9&150&1.357&1.17\\
\hline
\end{tabular}
\tablefoot{
$^{\rm a}$ Sky frames were taken 7 arcmin away from the object to ensure that there was no nebular contamination. 
}
\end{table}

\begin{figure*}
\centering
\includegraphics[width=18.5cm]{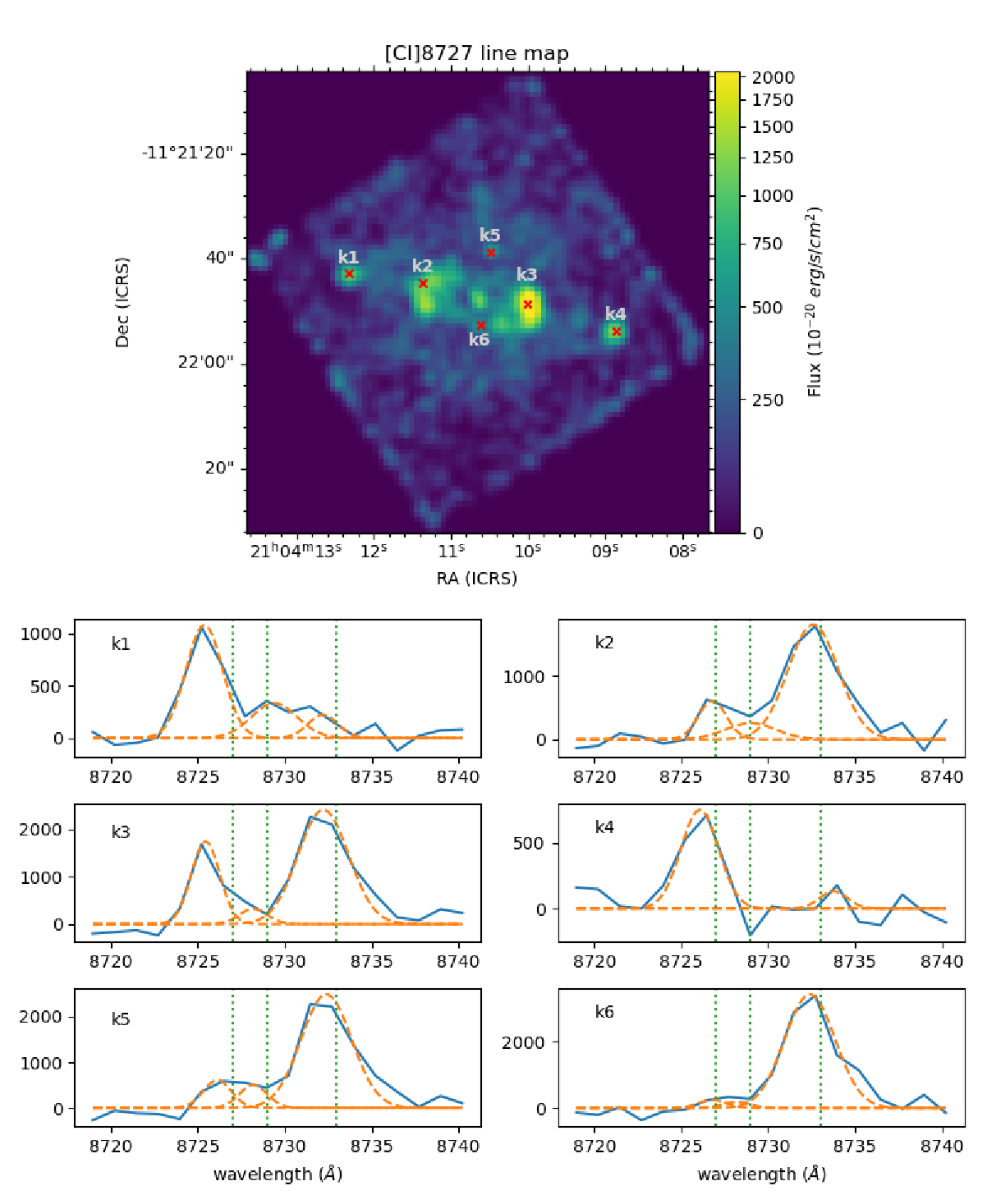}
\caption{Illustration of the spatial distribution of the atomic carbon that reveals the clumpy structures superimposed on the extended emission of NGC~7009. Upper panel: \carboni~$\lambda$8727 line MUSE flux map at a 5$\times$5 spatial binning. Six different clumpy structures on the nebula are indicated as red crosses and labeled from k1 to k6 \citep{Goncalves2004b}. The color bar corresponds to the intensity of the emission line. Bottom panels: Observed spectra (solid blue lines) in the range from 8720 to 8740~\AA~ extracted from the six distinct spaxels in units of 10$^{-20}$~erg~s$^{-1}$~cm$^{-1}$~\AA$^{-1}$. The three emission lines in this particular spectral range are \carboni~at~8727.1~\AA, [Fe~{\sc iii}] at~8728.8~\AA,~and He~{\sc i} at~8733.4~\AA, and they were fitted assuming a Gaussian profile (dashed orange lines). The vertical dashed green lines indicate the wavelengths in the rest frame.}
\label{figregions1}
\end{figure*}

\section{Near-infrared \carboni~$\lambda$8727 emission line in NGC~7009}

The fact that the outer pair of LISs in NGC~7009 shows enhanced \oxygeni~$\lambda$6300 and \nitrogena~$\lambda$5200 emission lines relative to the host nebula \citep[][]{Goncalves2003,Walsh2018,Akras2022}  in conjunction with H$_2$ emission \citep[][]{Akras2020b} motivated us to search in the Multi Unit Spectroscopic Explorer \citep[MUSE,][]{bacon2010S} integral field unit (IFU) data of NGC~7009 for the NIR \carboni~$\lambda$8727 line. A potential detection of the \carboni~line associated with the pair of knots would further support the scenario of photodominated knots, given that this line goes deeper into the molecular gas.

The MUSE IFU data cover an optical wavelength range from 480 to 930~nm\footnote{https://www.eso.org/sci/facilities/paranal/instruments/muse/inst.html}. This makes MUSE an ideal instrument for a search for the NIR \carboni~$\lambda$8727 line in extended sources such as PNe. In Fig.~\ref{figCI}, we present the first spatially resolved \carboni~$\lambda$8727 map of NGC~7009, extracted from the MUSE data obtained for the 097.D-0241(A) program (PI: R. L. M. Corradi)\footnote{We used the dataset of Corradi et al. instead of the science verification (SV) data \citep{Walsh2016,Walsh2018} because the latter were $\sim$65\% shallower. The larger seeing in the dataset of Corradi et al. (0.9-1.2 arcsec; see Table~\ref{Tab:Obs}) relative to the SV dataset (0.5-0.6 arcsec) has little impact on the current study of the LISs (d$\sim$3~arcsec) in NGC~7009. Spectroscopic analysis of NGC~7009 has shown that the two datasets agree well \citep{Akras2022}.}. The data acquisition and reduction for this program were detailed in \citet{GarciaRojas2022}. The sequence of target and sky observations, exposure times, air masses, and seeing conditions for NGC\,7009 are shown in Table~\ref{Tab:Obs}. The emission from the pairs of LISs as well as the inner ellipsoidal shell is readily discerned (Fig.~\ref{figCI}).

It should be noted, however, that two more lines lie close to the \carboni~line, namely the \ironiii~$\lambda$8729 and \helium~$\lambda$8733 lines. To verify the identification of the emission centered at 8727\AA~as the \carboni~line and exclude the possibility of a misidentification or a contribution from the other lines, three Gaussian components were considered for the fitting process.

The upper panel in Fig.~\ref{figregions1} displays the \carboni~$\lambda$8727 line map extracted from the datacube with a 5$\times$5 binning in the spatial dimensions (1.0$\times$1.0$''$ spaxels) to increase the signal-to-noise ratio of the emission lines. The lower panels show the spectra in the wavelength range from 8720\AA~to 8740\AA\ together with our best line fitting solution for six spaxels in the nebula, labeled k1 to k6 \citep{Goncalves2004}, which are indicated on the line map by red crosses. The two outer LISs (k1 and k4) clearly show an emission line centered at 8725--8726~\AA. The inner LISs k2 and k3 exhibit two emission lines centered at 8725--8726~\AA~and 8732--8733~\AA, respectively. At positions k5 and k6, we clearly obtain an emission line centered at $\sim$8733~\AA~and a weak emission line at $\sim$8726~\AA, which is slightly stronger in LIS k5. The difference in the line wavelengths is attributed to the low spectral resolution data and motion of the LISs.

According to this analysis, we argue that the two outer LISs (k1 and k4) are characterized by an intense \carboni~$\lambda$8727 emission, while the inner blobs (k2 and k3) exhibit a weaker \carboni~$\lambda$8727 emission accompanied by a stronger \helium~$\lambda$8733 emission. The k5 LIS that lies closer to the central star is dominated by the recombination \helium~emission, but a weak emission from atomic carbon is noticeable. Interestingly, the k6 region, which represents the spectrum of the ellipsoidal nebula, only shows strong \helium~$\lambda$8733 emission. This implies different physical conditions and ionization levels between the LISs and the host nebula \citep{Mari2023b}. The \helium~$\lambda$8733 emission line detected in the k2, k3, and k5 LISs is probably associated with the superposed emission from the host nebula. The \ironiii~$\lambda$8729 line is only marginally found in k1, making its detection uncertain. It is worth mentioning, however, that the singly ionized lines \ironii~$\lambda$8617 (Bouvis et al. in preparation) and \ironii~1.644$\mu$m \citep{akras2024accepted} have also been detected in LIS k1.

\begin{table}
\caption{Optical emission line fluxes from specific pseudo-slits or nebular regions$^{\rm a}$.} \label{tab:fluxlinesPN}
\centering
\renewcommand{\arraystretch}{1.2} % Default value: 1
\begin{tabular}{lllllll}
\hline

Ion           &I$_{F}$ &I$_{S_F}$ & I$_{S_{k1}}$ & I$_{S_{k2}}$ & I$_{S_{k4}}$ \\
\hline
He~{\sc ii}~4686~\AA\ & 5.44 & 17.45  & 0.692   & 9.78     & 0.652 \\
H~{\sc i}~4861~\AA\   & 100        & 100    & 100     & 100    & 100   \\ 
\oiiia                & 388    & 391    & 398     & 399    & 449   \\ 
\nia                  & 0.09   & 0.087  & 3.577   & 0.104  & 4.135  \\
He~{\sc ii}~5412~\AA\ &  1.32  & 1.345  & 0.196   & 0.758  & 0.292 \\
%\cliiia               &  0.435 &  0.5  & 0.454   & 0.860  & 0.932 \\
%\cliiib               &  0.539 &  0.6  & 0.546   & 0.796  & 0.797 \\
\niia                 & 0.395  &  0.408 & 4.388   & 0.582  & 5.143  \\
He~{\sc i}~5876~\AA\  & 14.42  &  15.35 & 15.72   & 16.09  & 14.96  \\
\oi                   & 0.559 &  0.544 & 21.7    & 0.746  & 19.72  \\
%\siiia                &  1.39 &  1.4   & 1.39    & 3.07   & 3.39 \\
%\niic                 &  4.79 &  -     & 5.63    & 108    & 110  \\
H~{\sc i}~6563~\AA\   & 266    &  288   & 289     & 288    & 289 \\
\niib                 & 15.64  & 18.10  & 253     & 33.1   & 307   \\
He~{\sc i}~6678~\AA\  &  3.68  &  4.12  & 4.352   & 4.393  & 4.105  \\
\siia                 &  1.377 &  1.71  & 26.40   & 3.415  & 34.77 \\ 
\siib                 &  2.284 &  2.89  & 37.06   & 5.777  & 43.27 \\
%\ariii                & 15.7   &  -    & 14.9    & 27.4   & 29.2  \\
%\oiia                 &  1.36  & 1.3:  &  1.19   & 8.56   & 7.45   \\
%\oiib                 &  1.17  & 1.1   &  1.11   & 7.19   & 6.36  \\
%\siiib                & 24.5   & 25.2  & 23.2    & 48.7   & 51.3  \\
\carboni~8727~\AA\    &   -    &  0.011 &  0.172  & 0.012  & 0.170  \\
\ironiii~8729~\AA\   &  0.018 &  0.005 &  0.039  & 0.004  & 0.041  \\
He~{\sc i}~8733~\AA\  &  0.075 &  0.024 &  0.046  & 0.025  & 0.069  \\
\hline
\vspace{0.20cm}
F(\hb)(10$^{-13}$)    & -      & 208   &  1.25   & 35.8   & 0.65   \\
\vspace{0.20cm}
%{\displaystyle \frac{\rm{He~II~\lambda4686}}{\rm{He~I~\lambda8733}}}$ & 72    & 727     &  15    & 391   & 384  & 9.5   \\
%\vspace{0.20cm}
%${\displaystyle \frac{\rm{He~II~\lambda5412}}{\rm{He~I~\lambda8733}}}$ & 17.6  & 56      &  4.26  & 30   & 30   & 4.23  \\
%\vspace{0.20cm}
%${\displaystyle \frac{\rm{He~I~\lambda5876}}{\rm{He~I~\lambda8733}}}$  & 195    & 640     &  340   & 643   & 684   & 217  \\
${\displaystyle \frac{\rm{[C~I]~\lambda8727}}{\rm{H\,\alpha}}}$~$^\dag$ & (6.77)$^{\dag\dag}$    & 3.8     &  59.5    & 4.2   & 58.8   \\
\hline        
\hline
\end{tabular}
\tablefoot{
$^{\rm a}$I is the intensity of the lines on the scale H$\beta$ = 100. The indices S and F refer to the results obtained with {\sc satellite} \citep{Akras2022} and \citep{Fang2011}. The labels k1, k2, and k4 correspond to the nomenclature given in \cite{Goncalves2004b}. $^\dag$(10$^{-5}$). $^{\dag\dag}$Assuming that \ironiii~8729~\AA\ corresponds to \carboni~8727~\AA.}
\end{table}

\cite{Fang2011} presented a very deep spectrum of NGC~7009 covering a wavelength range from 3040 to 11000~\AA, with almost 1200 detected and identified emission lines. Several recombination lines of C$^+$ ions were identified, as well as the two NIR forbidden \carboni~lines at 9824 and 9850~\AA. The authors did not report the detection of the \carboni~$\lambda$8727 line, however. In contrast, the same authors reported the identification of the \ironiii~$\lambda$8729 and \helium~$\lambda$8733 lines. It is worth mentioning that the slit only covered the ellipsoidal inner shell of the nebula, centered 2-3 arcsec from the central star in the southern direction. Hence, the identification of the \helium~$\lambda$8733 by \cite{Fang2011} agrees with our findings in the k6 region, while the report of the \ironiii~$\lambda$8729 line may be a misidentification of the \carboni~$\lambda$8727 line in the k2 and k3 LISs. 

The absence of the \carboni~$\lambda$8727 line in the main nebula (k6) in conjunction with the detection of the \carboni~$\lambda\lambda$9824,9850 by \cite{Fang2011} implies a high \carboni~ratio. For T$_{\rm e}\sim$10000~K
and 5000<n$_{\rm e}$<10000~cm$^{-3}$, both mechanisms are valid (the recombination of C$^{+}$ or the collisional excitation by free electrons). In the case of 1000<n$_{\rm e}$<5000~cm$^{-3}$, collisional excitation of atomic carbon is the most probable mechanism \citep[see fig.~2 in][]{Liu1995CI}. 
The high observed n$_{\rm e}$ (between 3500~cm$^{-3}$ and 10000~cm$^{-3}$) reported for the main elliptical shell of NGC~7009 based on the \sulfurt~$\lambda$6716/$\lambda$6731 diagnostic line ratio \citep[e.g.][]{Goncalves2003,Goncalves2006,Fang2011,Walsh2018,Akras2022,SeongJae2022,Hyung2023}, indicates collisional excitation of C.

In Table~\ref{tab:fluxlinesPN}, we list the intensities of some typical emission lines computed for the outer LISs, the inner blobs, and Fang's slit from the MUSE data using the specific slit module in the code called Spectroscopic analysis tool for intEgraL fieLd unIt daTacubEs {\sc satellite} \citep{Akras2022}. {\sc satellite} simulates the long-slit spectrum of \cite{Fang2011} defining a pseudo-slit with similar sizes and placed at approximately the same position on the nebula. The intensity of the \ironiii~$\lambda$8729 line measured by \cite{Fang2011} is comparable to the sum of the \carboni~$\lambda$8727 and \ironiii~$\lambda$8729 lines for the same region (second and third columns of Table~\ref{tab:fluxlinesPN}). When we combine the intensity of the \carboni~$\lambda$8727 line from the MUSE data (third column in Table~\ref{tab:fluxlinesPN}) with the \carboni~$\lambda$9824 and $\lambda$9850 line intensities from \cite{Fang2011}, we obtain R(\carboni)$\sim$2.2, which implies collisional excitation of C$^0$ in high electron density gas (n$_{\rm e}$=6-8$\times$10$^4$~cm$^{-3}$). This is orders of magnitude higher than the observed value (n$_{\rm e}$ =1300-1500~cm$^{-3}$; \cite{Goncalves2003,Walsh2018,Akras2022}). We thus argue that the \carboni~$\lambda\lambda$9824,9850 emission lines more likely originate from the inner and outer LISs.

\begin{figure}
\centering
\includegraphics[width=9cm]{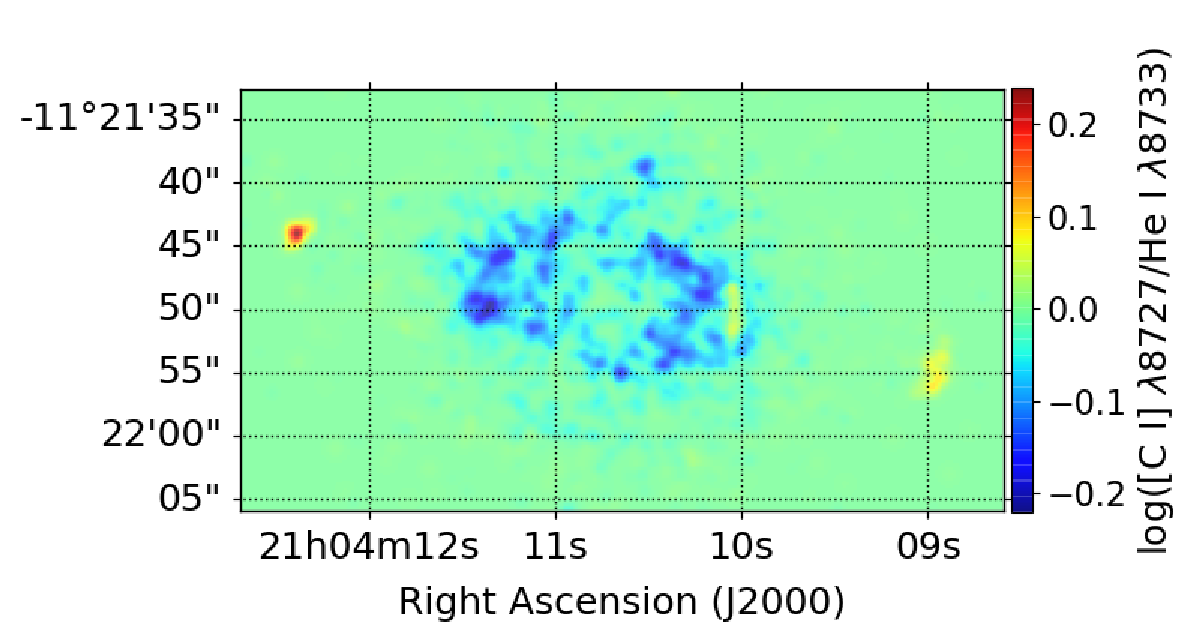}
\caption{\carboni~$\lambda$8727/\helium$~\lambda$8733 line ratio map from MUSE data of NGC~7009.}.
\label{figCINGC7009}
\end{figure}

\begin{figure*}
\centering
\includegraphics[width=14cm]{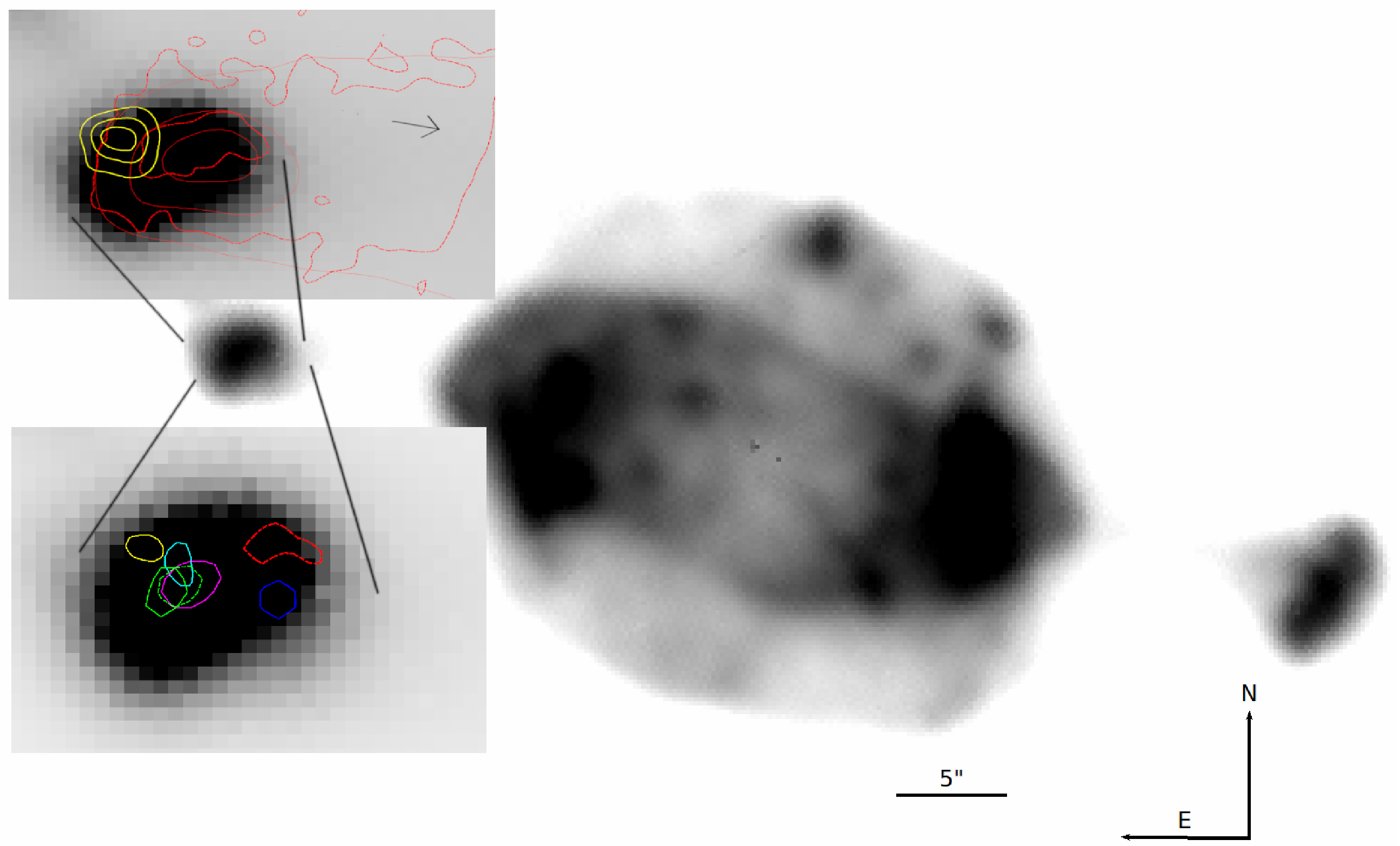}
\caption{Radial ionization stratification in the LISs of NGC~7009. Isocontours normalized to the maximum value of the emission lines at the position of knot k1, overlaid on the MUSE \nitrogen~$\lambda$6584 emission line map from \cite{Akras2022}. Upper left panel: H$\alpha$ (red), Br$\gamma$ (dashed red) and H$_2$ (yellow) emission lines. Lower left panel: Br$\gamma$ and H$_2$ (dashed red and yellow; \citealt{Akras2020b}), \ironii~(cyan; \citealt{akras2024accepted}), \oxygeni, \sulfurt~and \oxygeniii~(dashed green, magenta and blue, respectively; \citealt{Akras2022}) and \carboni~(green, this work). The field of view of the upper and lower panels is 7.5$\times$6~arcsec and 6.5$\times$4.8~arcsec, respectively. The arrow indicates the direction of the central star. North is up and east is to the left.}
\label{contours}
\end{figure*}

The intensity of the \carboni~$\lambda$8727 line in the outer LISs (Cols. 4 and 6) is higher by a factor of 15 (relative to the host nebula), similarly to other low-ionization lines (e.g., \nitrogen~$\lambda$6584, \sulfurt~$\lambda$6716, $\lambda$6731, and \oxygeni~$\lambda$6300). When we assume that the \carboni~$\lambda$9824 and $\lambda$9850 lines come only from the other LISs, R(\carboni) becomes larger (highly uncertain, but close to 10). This is consistent with the recombination of singly ionized carbon or collision excitation in a gas with n$_{\rm e}$ between 10$^{3}$ and 10$^{4}$~cm$^{-3}$. 

In Fig.~\ref{figCINGC7009}, we present the spatial distribution of the \carboni~$\lambda$8727/\helium$~\lambda$8733 line ratio from the MUSE data of NGC~7009 considering a signal-to-noise ratio $>$3. This map clearly shows that the central part of NGC~7009 is dominated by an intense \helium$~\lambda$8733 emission, while the outer LISs exhibit an enhanced \carboni~$\lambda$8727 line emission.

\section{Discussion}

By definition, LISs are characterized by enhanced low-ionization emission lines (relative to \ha) compared to the surrounding nebular gas. The detection of the H$_2$ 2.12~$\mu$m emission in LISs has unveiled the following structure: A thin layer of partially ionized gas that engulfs the molecular core. 

The emission line stratification of the LISs in NGC~7009 has been demonstrated through the radial analysis of the nebula using the {\sc satellite} code \citep[see fig.~7 and table~2 in][]{Akras2022}. Low-ionization lines peak at 1~arcsec farther from the central star than the moderate- to high-ionization lines.

Figure~\ref{contours} displays the isocontours, normalized to the maximum intensity, of several emission lines found to emanate from knot k1, overlaid on the \nitrogen~$\lambda$6584 line map from MUSE. The bottom left panel shows the distribution of the H$\alpha$ (solid red), Br$\gamma$ (dashed red), and H$_2$ (yellow) emission lines obtained from \cite{Akras2020b}. The H$\alpha$ and Br$\gamma$ lines are cospatial, as expected, and extend in the western direction. The spatial offset between the H$\alpha$ and Br$\gamma$ emission with the H$_2$ emission is easily discerned. The lower right panel shows the isocontours of the maximum intensity of more lines. The stratification of the emission lines is evident. The \oxygeniii~emission peaks closer to the central star, followed by the \sulfurt~and \nitrogen~emission, the cospatial \carboni, \oxygeni, \ironii~emission, and finally, the H$_2$ emission at a larger distance from the central star.

\begin{figure}
\centering
\includegraphics[width=7.75cm]{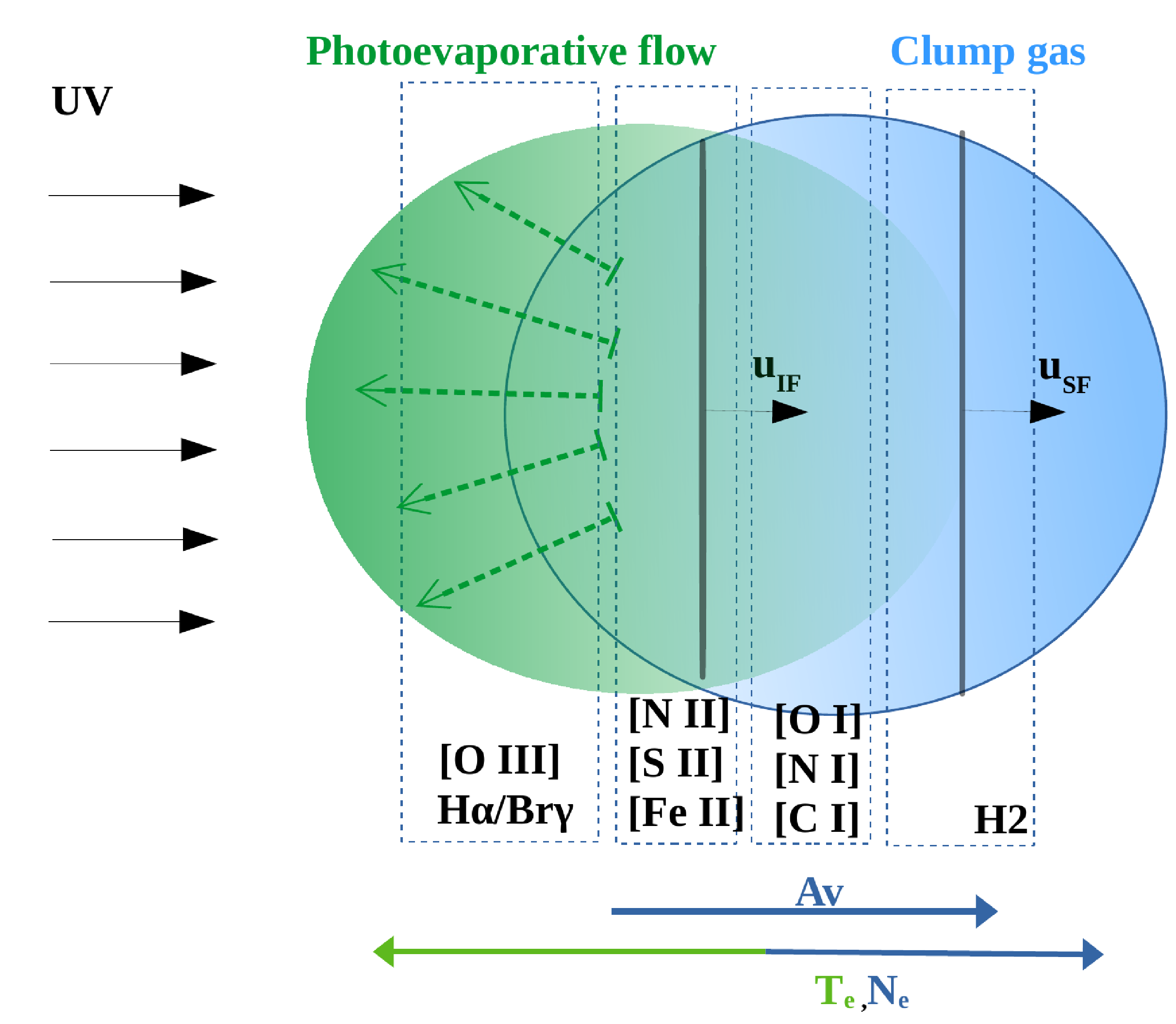}
\includegraphics[width=8.25cm]{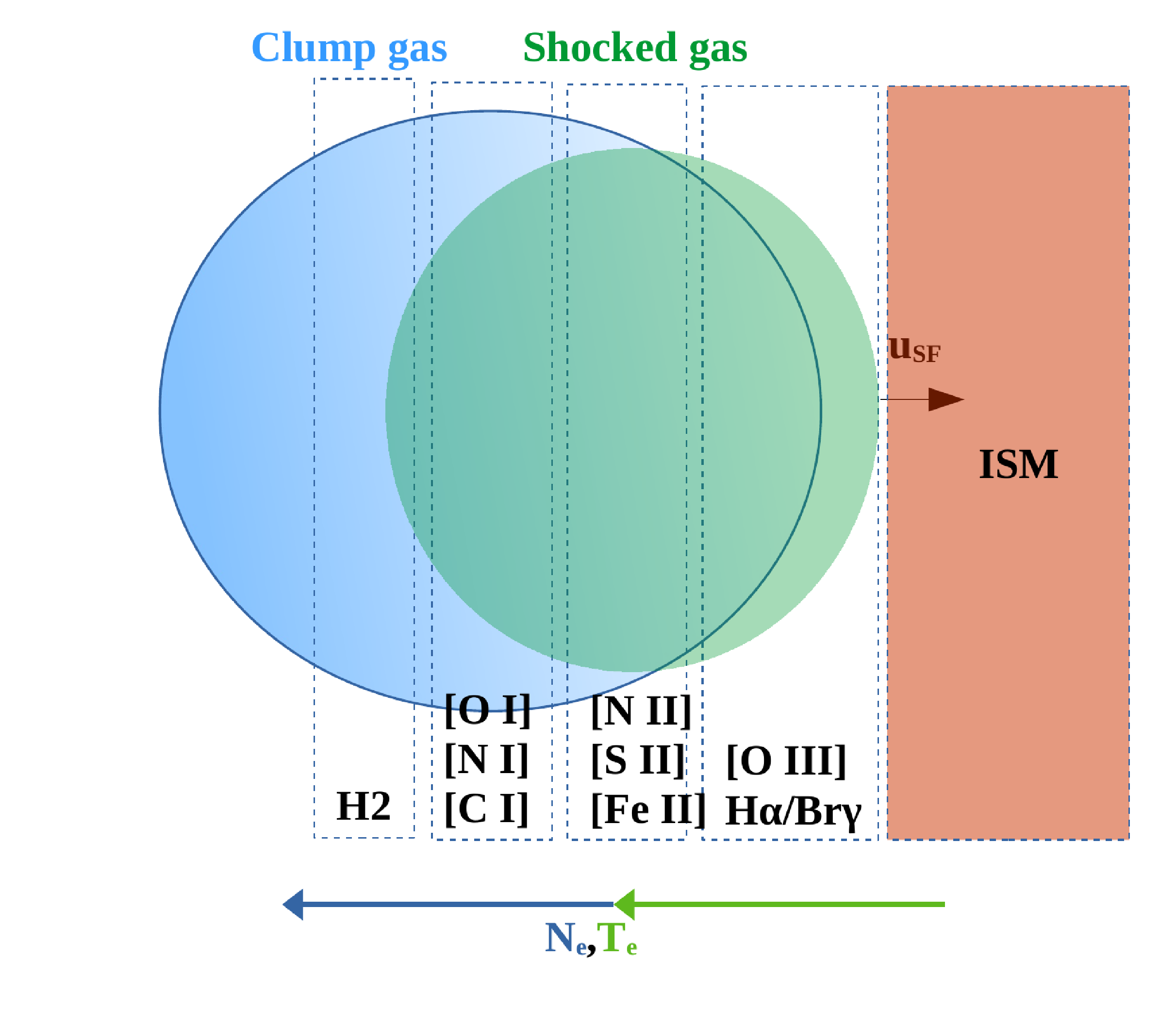}
\caption{Schematic representation of the emission line stratification in a photoevaporated (upper panel) and shock-heated due to the interaction with nebular gas or ISM (lower panel) clump. The dashed green arrows and green region represent the photoevaporative flow due to the illumination of the clump (blue region) by the UV stellar radiation field (black arrows) and the shock-heated gas by the reverse shock propagating into the clump as well as the shock-heated gas in case of an interaction with the surrounding medium due to the outward motion of the clump. The solid blue and green arrows at the bottom of each cartoon show the increase in the extinction Av, $T_{\rm e}$, and $n_{\rm e}$, respectively. u$_{IF}$ and u$_{SF}$ correspond to the velocity of the ionization and shock fronts, respectively.}
\label{figCIregions1}
\end{figure}

This emission line stratification is expected for a clump illuminated by an external radiation source (see a scheme of this scenario in the upper panel in Fig.~\ref{figCIregions1}), but it contradicts the scenario of shock-excited bullets that move outward and interact with the AGB remnant or the interstellar medium (lower panel in Fig.~\ref{figCIregions1}). In the latter case, the stratification of the emission lines should be in the opposite direction. In particular, the high- to moderate-ionization emission should lie just behind the shock front, followed by the low-ionization emission, as the temperature of the gas decreases with the distance from the shock front. 
The schematic cartoons in Fig.~\ref{figCIregions1} illustrate the case of a UV-illuminated (photoevaporation) and a shock-excited clump. Representative examples of the emission line stratification in a shock-excited gas are also provided by \cite{Bocchino2000} and \cite{Alarie2019}.

Shocks are usually unveiled by enhanced \oxygeniii/\ha~ line ratios, which reflect the increase in T$_{\rm e}$. \cite{Guerrero2013} discussed the enhancement of the \oxygeniii/\ha~ratio in PNe and pointed out that it is more frequent in bipolar outflows, expanding shells, and clumps. Particularly for NGC~7009, an enhancement of the \oxygeniii/\ha~ratio as well as T$_{\rm e}$\sulfur~has been observed at the outer edges of the LISs from the HST line images \citep[][]{Guerrero2013} and in MUSE line maps \citep[][]{Walsh2018,Akras2022}.

The radial analysis of NGC~7009 with the {\sc satellite} code also revealed an increase in $T_{\rm e}$([S~{\sc iii}]) by approximately 1000~K at the position of LISs, where the \oxygeniii/\ha~ratio becomes higher. This increase in the \oxygeniii/\ha~ratio might be associated with a shock-heated gas. We recall that the emission of molecular hydrogen in the LISs of NGC~7009 has been attributed to the UV-pumping process in highly dense gas \citep[$>$10$^4$~cm$^{-3}$,][]{Akras2020b}, and high densities like this have been reported by \cite{SeongJae2022,Hyung2023}, but shocks have not yet been ruled out. Shocks contribute little to the total H$_2$ emission compared to the UV \citep{Natta1998}.

Shock interaction is not the only process that can lead to the heating of the gas. In PDRs and PNe with a significant number of small dust grains, the effect of the photoelectric heating by dust grains becomes very strong and may cause the increase in the \oxygeniii/\ha~ratio \citep{Hollenbach1991,dopita2000,Stasinska2001}. According to these studies, photoelectric heating can provide a natural explanation for several problems: (i) the temperature gradient in PNe, (ii) the discrepancy in the \oxygeni~$\lambda$6300 between the observations and models, and (iii) the systematically lower $T_{\rm e}$ obtained from the Balmer jump, and other diagnostics. Molecular hydrogen in LISs also implies the presence of dust. UV shielding by dust may be significant, but more importantly, dust is needed to form H$_2$.

Near-infrared neutral \carboni~lines are important coolants in PDRs \citep{Burton1992}, and they define the transition zone between the ionized and neutral gas \citep{Tielens1985a,Tielens1985b,Hollenbach1999}. The detection of the \carboni~$\lambda$8727 line, which we reported here for the first time in LISs, has added a new piece to the puzzling problem of their origin. The question is whether the \carboni~emission indicates that a considerable amount of hidden atomic carbon is stored in these microstructures, or if it is associated with the dissociation of CO (Gon\c{c}alves et al. in preparation).

Considering the intense UV radiation field from the central stars in PNe, the photoevaporation of dense molecular clumps \citep{Mellema1998} is possible. The stellar UV radiation illuminates the inner part of the dense clumps and results in the formation of a photoevaporating flow that moves backward. The moderate ionization lines emerge from this particular gas and display a peak closer to the central star, while the low-ionization lines originate from a region closer to the molecular core (at a larger distance from the central star) where the UV field is attenuated. The spatial distributions of the \oxygeni~6300/\ha~and \nitrogena~5200/\ha~ line ratios at the outer LISs of NGC~7009 exhibit higher values deeper in their cores \citep{Akras2022a}. In addition to the optical emission, the H$_2$ emission from the outer LISs peaks at an even larger distance than the low-ionization lines (see Fig.~\ref{contours}).

The estimated spatial offset between the H$_2$ and Br$\gamma$ line emissions from the LISs in NGC~7009 is $\sim$1.2~arcsec, which corresponds to $\sim$2.0$\times$10$^{16}$~cm adopting a distance of 1180~pc \citep{BailerJones2021} based on the {\it Gaia} EDR3 parallaxes \citep[][]{Gaia2021}. A comparable spatial offset between the peaks of the H$_2$ 1-0 and Br$\gamma$ lines of about 10$^{16}$~cm has also been reported in molecular clouds of star-forming regions and was attributed to the photoevaporation of gas \citep[e.g. IC~1396, Cygnus OB2, and the Carina nebula;][]{Hartigan2015,Carlsten2018}.

According to the photoevaporation model of \citet[][]{Carlsten2018}, the offset between the H$_2$ 1-0 and Br$\gamma$ line becomes smaller when the flux of ionizing photons increases or the illuminated clumps become smaller. For the central star of NGC~7009~($T_{\rm eff}$ = 80\,000-90\,000~K and log(L/L$_{\odot}$) = 3.5-3.7\footnote{These luminosities were estimated considering distances of 1.4~kpc \cite{Sabbadin2004} and 0.86~kpc \cite{Goncalves2006}. Considering the geometric distance of 1.18~kpc (1.13 and 1.26~kpc are the 16\% and 84\% percentiles) in the {\it Gaia} Early Data Release 3 \citep{BailerJones2021}, we get log(L/L$_{\odot}$) = 3.55 and 3.8, respectively.}; \cite{Sabbadin2004,Goncalves2006}), the flux of the incident ionizing photons, $F$ (photons~cm$^{-2}$~s$^{-1}$), at the surface of the k1 knot with a radius of r$_c$=2$\times$10$^{16}$~cm \citep{Akras2020c} and at a distance of r=3.8$\times$10$^{17}$~cm (22.6~arcsec) from the central star, is log($F$)$\sim$18. Although the photoevaporation model is limited to log($F$)=13.25 and the clump radius is limited to r$_c$ $\geq$10$^{17}$~cm (see fig.17 in \citealt{Carlsten2018}), we claim that the modeled offset for the parameters of NGC~7009 could be as low as 10$^{16}$~cm, which is comparable with the observations.

The spatial offset between the \sulfurt~and \ha~line emissions at knot k1 has also been computed as $\sim$2.0$\times$10$^{16}$~cm \citep[$\sim$1.2~arcsec, see Fig.~\ref{contours} and][]{Akras2022}. \cite{Hartigan2015} has measured comparable spatial offsets between the \sulfurt~and \ha~line emissions in molecular clouds of star-forming regions. In both cases, the \ha~line displays a peak closer to the UV source.

According to the photoevaporation scenario, the presence of atomic carbon in gaseous state in clumps may imply the presence of the CO molecule, which is being dissociated by the UV radiation from the central star. However, CO emission has not yet been found in LISs (Gon\c{c}alves et al. in preparation), except for the cometary knots in the Helix \citep{Huggins1992,Hugginsetal2002,Andriantsaralaza2020}. 

Another possible mechanism that can also lead to the emission of neutral carbon in gaseous state is shock waves.  Evidence for shocks (either J- or C-type) have been found in Herbig-Haro (HH) objects through the detection of the H$_2$, \ironii~1.644~$\mu$m and \carboni~$\lambda\lambda$9830,9850 emission lines \citep[][]{Smith1994,Smith2003,Mccoey2004}. To account for the emission of \ironii~and \carboni~lines in HH objects, \cite{Mccoey2004} assumed that Fe and C are first released from the dust grains into the gas phase by the passage of a C-type shock with a velocity between 25 and 45~km~s$^{-1}$ in high-density (10$^4$~cm$^{-3}$) gas \citep{May2000}. The \carboni~$\lambda$8727/\ironii~1.644~$\mu$m and \carboni~$\lambda$8727/H$_2$~2.12~$\mu$m line ratios for the k1 LIS in NGC~7009 were computed to be 0.125 and 0.05, respectively \cite{Akras2020b}, and they were found to be comparable with the observed and modeled ratio for HH99 \citep{Mccoey2004}. However, quantitative PDR and shock models that provide predictions for the intensities of the  \ironii~1.644~$\mu$m, H$_2$~2.12~$\mu$m, and\carboni~$\lambda\lambda$8727,9850, among other low-ionization lines, are needed for the proper investigation of k1 and other LISs. Because of the outward motion of a clump and its interaction with the AGB envelope or interstellar medium (ISM), shock-heated gas does not result in the observed emission line stratification.

It is important to point out that low-velocity shock waves are also present in the photoevaporation process.  As the photoevaporating flow moves backward with respect to the direction of the k1 LIS, it interacts with the stellar wind and forms a forward- and a reverse-shock wave \citep[see e.g.][]{Mellema1998,Bron2018}. This particular reverse shock sweeps up the material in the LIS or clump, and it might cause the release of C and Fe from the dust grains into the gas phase. Shock waves with velocities $<$50~km~s$^{-1}$, depending on dust grain size and composition, are sufficient for grain destruction to some extent and for the release of elements into the gas phase\ \citep[e.g.][]{Jones1994,May2000}. 

\section{Conclusions}
We presented the first spatially resolved \carboni~$\lambda$8727 emission line map of the planetary nebula NGC~7009 extracted from the MUSE datacube. \carboni~$\lambda$8727 emission is mainly detected at LIS pairs. The identification of the emission line was verified by comparing the spectra in six different regions. The outer LIS pair is characterized by a strong \carboni~$\lambda$8727 line and a three times fainter \helium~$\lambda$8733, while the core nebula is mainly characterized by the \helium~$\lambda$8733 line. We cannot determine the mechanism (collision or recombination) through which the \carboni~line is emitted based on the data we used.

Nevertheless, the detection of the forbidden \carboni~$\lambda$8727 line in the LISs, an important coolant in PDRs, provides additional evidence to support the scenario in which LISs are low-ionization dense molecular structures surrounded by a partially ionized or neutral gas. The observed stratification in the LISs of NGC~7009 disagrees with the result of shock interaction with the surrounding medium, but it is consistent with the outcome from the photoevaporation of a molecular clump illuminated by the UV radiation from the central star. The interaction of the inward photoevaporated gas with the outward stellar wind may result in a weak reverse-shock wave that causes the enhanced emission from low-ionization species. High spatial resolution observations are required to enrich our understanding of LISs in PNe.

\begin{acknowledgements}
We would like to thank the anonymous referee for his/her detailed report on the manuscript, which indeed helped us to significantly improve the paper. The research project is implemented in the framework of H.F.R.I call \lq\lq~Basic research financing (Horizontal support of all Sciences)\rq\rq~under the National Recovery and Resilience Plan \lq\lq Greece 2.0\rq\rq~funded by the European Union – NextGenerationEU (H.F.R.I. Project Number: 15665). SA and LK acknowledge support from H.F.R.I. DRG acknowledges FAPERJ (E-26/211.370/2021; E-26/211.527/2023) and CNPq (403011/2022-1; 315307/2023-4) grants.  JG-R acknowledges financial support from the Spanish Ministry of Science and Innovation (MICINN) through the Spanish State Research Agency, under Severo Ochoa Centres of Excellence Programme 2020-2023 (CEX2019-000920-S). The authors would like to acknowledge Dr. Henri Boffin for the reduction of the MUSE data. The following software packages in Python were used: Matplotlib \citep{Hunter2007}, NumPy \citep{Walt2011}, SciPy \citep{SciPy2020} and AstroPy Python \citep{Astropy2013,Astropy2018}.

\end{acknowledgements} 

\section*{Data availability}
The data are available upon request to the corresponding author. The observations are also available in the ESO Archive.

%\PassOptionsToPackage{unicode}{hyperref}
\bibliographystyle{aa}
\bibliography{references}

\end{document}